\documentclass[aps,prl,showpacs,twocolumn,floatfix]{revtex4}
\usepackage{graphics,graphicx}
\usepackage{bm,bbm}
\usepackage{amssymb,amsmath,amsfonts}
\usepackage{mathrsfs}
\usepackage{calc}

\def\prb#1#2#3{Phys.~Rev.~B~{\bf #1},\ #2\ (#3)}

\def\prl#1#2#3{Phys.~Rev.~Lett.~{\bf #1},\ #2\ (#3)}

\def\etal{{\it et al.}}

\newcommand{\setfigwidthratio}[1]{\columnwidth*\real{ #1 }}
\newcommand{\bra}[1]{\ensuremath{\langle #1 |}}
\newcommand{\ket}[1]{\ensuremath{| #1 \rangle}}
\newcommand{\bracket}[2]{\ensuremath{\langle #1 | #2 \rangle}}

\def\half{\mbox{$1\over2$}}
\def\halfi{\mbox{$i\over2$}}
\def\bpmat{\begin{pmatrix}}
\def\epmat{\end{pmatrix}}
\def\bmat{\begin{matrix}}
\def\emat{\end{matrix}}
\def\sP{{\mathscr{P}}}
\def\RR{\mathbb{R}}
\def\bT{{\bf T}}
\def\bR{{\bf R}}
\def\1{\mbox{1\hskip-.25em l}}

\newcommand{\SA}{{S\!/\!A}}
\newcommand{\LR}{{L\!/\!R}}
\newcommand{\qb}{{q_\mathrm{b}}}
\newcommand{\qi}{{q_\mathrm{i}}}

\begin{document}

\title{\bf Controlling Fano Profiles via Conical Intersections}
\author{Yoav Berlatzky} \email{yoavbe@tx.technion.ac.il}
\affiliation{Department of Physics, Technion -- Israel Institute of
  Technology, Haifa 32000, Israel} \author{Shachar Klaiman}
\email{shachark@tx.technion.ac.il} \affiliation{Shulich Department of
  Chemistry and Minerva Center for Nonlinear Physics of Complex
  Systems, Technion -- Israel Institute of Technology, Haifa 32000,
  Israel.}

\begin{abstract}
  An \textit{ab initio} analytical model for Fano resonances in
  two-channel systems is presented. We find that the lineshape
  parameter $q$ factors into a background contribution $\qb$ that
  depends only on the uncoupled channels and an interaction
  contribution $\qi$ that is affected by the coupling between the
  channels, revealing how the overall lineshape parameter $q$ may be
  controlled. In particular, we show how conical intersections of the
  background phase shifts have an important role in the interplay
  between $\qb$ and $\qi$.  Finally, control of Fano transmission
  profiles through $\qb$ and $\qi$ is demonstrated for quantum
  billiards.
\end{abstract}
\pacs{73.63.Kv, 33.40.+f, 73.23.$-$b, 72.15.Qm}
\maketitle


Fano resonances are ubiquitous whenever discrete levels interact with
a continuum. They have been observed in many areas of physics: atomic
and molecular systems \cite{fano1,hossein}, quantum dots
\cite{QDfano}, carbon nanotubes \cite{kim}, and microwave billiards
\cite{srotter}. When probed, Fano resonances yield their signature
scattering cross-section profile \cite{fano1}
\begin{equation}
  \label{eq:fano}
  \sigma\propto\frac{\left(\epsilon+q\right)^2}{1+\epsilon^2}\,, 
\end{equation}
which has been used as a fitting function for the results of many
experiments as well as numerical calculations. Here $\epsilon$ is the
dimensionless reduced energy (which is measured in units of the
resonance's half-width and shifted so that it vanishes at the position
of the resonance), while $q$ is the Fano lineshape asymmetry
parameter.  The latter is proportional to the ratio between the
resonant and non-resonant transition amplitudes \cite{fano1}.
Unfortunately, this definition does not reveal any means of profile
control. In this work we suggest how to manipulate the asymmetry
parameter $q$ through a novel general analytical model for
two-channels systems that can be specifically applied to two
dimensional open quantum billiards (QBs).

\begin{figure}[htbp]
  \includegraphics[width=\setfigwidthratio{1}]{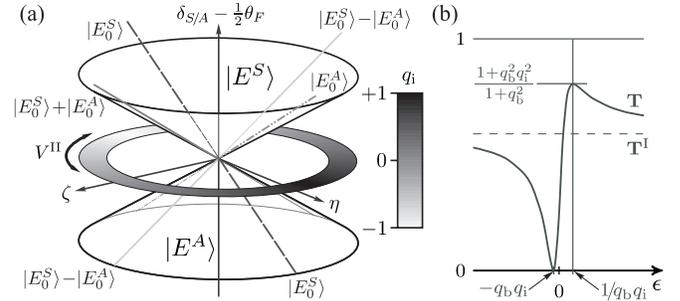}
  \caption{(a) Parameter space for controlling $q$ near a conical
    intersection of the background phase shifts. The upper cone
    corresponds to the state $\ket{E^S}$ and its phase shift
    $\delta_S$ (measured relative to $\half \theta_F$). To leading
    order, this state is an azimuth-dependent linear combination of
    the vertex states $\ket{E^\SA_0}$. Changing the coupling
    $V^\mathrm{II}$ corresponds to rotating the $\qi$ ``compass
    wheel''. (b) The general form of the transmission Fano lineshape.
    The transmission zero is obtained at $\epsilon=-\qb\qi$, while the
    maximal transmission value $(1+\qb^2\qi^2)/(1+\qb^2)$ is attained
    at $1/\qb\qi$. The transmission approaches the background value
    $\bT^\textrm{I}$ as $|\epsilon|\to\infty$. }
  \label{fig:twistcone}
\end{figure}

Our analysis reproduces the famous result of Eq. \ref{eq:fano}.  However,
the Fano lineshape parameter $q$ now factors as:
\begin{equation}
  \label{eq:qfact}
  q=\qb\cdot \qi.
\end{equation}
Here $\qb$ is responsible for the contribution of the uncoupled lower
(background) channel, whereas $\qi$ reflects the effect of the
interaction between the two channels on the resulting profile.  Our
results show the crucial role played by the background transmission
unity peaks, which correspond to conical intersections of the
background phase shifts, see Fig.~\ref{fig:twistcone}(a).  The factor
$\qb$ corresponds to the radial distance from the vertex, and $\qi$
behaves like the cosine of the azimuthal angle.  Finally, the
factorization opens the door to resonance profile control which we
demonstrate for a quantum billiard model. This elucidates Fano
$q$-reversals in a manner which is more complete and general than
previous attempts \cite{ab_qd_fano_theory}.

\textit{Two-channel Model.}--- In order to analyze a multi-dimensional
scattering process that yields a Fano profile, it is often useful to
represent the system as a set (\textit{albeit} infinite) of coupled
channels. We assume a single-particle time-reversal invariant (TRI)
multichannel system where only the lowest channel is open and the
scattering energy is in the vicinity of an isolated bound state in the
second channel, with other bound states energetically separate so that
we may disregard them.  The two-channel Hamiltonian is partitioned as
$H=H^0+V^\mathrm{I}+V^\mathrm{II}$, so that the ``distorted wave''
formalism \cite{taylor}, which treats scattering as a conceptual
two-stage process, may be applied. The first scattering stage, with
S-matrix $S^\mathrm{I}$, describes scattering due to $V^\mathrm{I}$
relative to the $H^0$ states in the first uncoupled channel. The
potential $V^\mathrm{I}$ also accounts for the bound state
$\ket\varphi$ at energy $E_\varphi$ in the second channel. The second
scattering stage's $S^\mathrm{II}$ describes the scattering of the
$H^0+V^\mathrm{I}$ states under the influence of inter-channel
interaction $V^\mathrm{II}$, which turns the bound state into a
resonance. The scattering matrix for the entire two-stage process is
$S=S^\mathrm{I}S^\mathrm{II}$ \cite{taylor,MSP}. In the present case
it is assumed that $S^\mathrm{I}$ is known, and we shall show how
$S^\mathrm{II}$ is approximated using the generic Fano model
\cite{fano1}.

The first stage is a simple TRI 1D problem. Its S-matrix may be
written as
\begin{equation}
  \label{eq:1DTRISmatrix}
  S^{\mathrm{I}}=\bpmat & t^{\phantom-} && r^- \\ & r^+ && t^{\phantom+} \epmat 
  =\bpmat & t^{\phantom-} && re^{-i\delta_r} \\ & re^{+i\delta_r} && t^{\phantom+} \epmat \bmat \\. \emat
\end{equation}
This matrix has eigenvalues that satisfy $e^{2i\delta_S}=t+r$ and
$e^{2i\delta_A}=t-r$, hence the notations $\delta_{\SA}$ refer to the
symmetric/antisymmetric combinations of $t$ and $r$. The ambiguity in
the above decomposition may be removed by one of the discrete gauge
choices $-\pi/2 < \delta_r \le \pi/2$ or $\delta_S\ge\delta_A$.  We
also define the Friedel angle $\theta_F=\delta_S+\delta_A$
\cite{friedel}, the difference $\Delta\theta=\delta_S-\delta_A$, and
the states
\begin{subequations}
  \label{eq:EigenStates}
  \begin{align}
    \ket{E^S}&
    =\frac{e^{-i\delta_S}}{\sqrt 2}\left[e^{-\halfi \delta_r}\ket{E^+} + e^{+\halfi \delta_r}\ket{E^-} \right] \\
    \ket{E^A}&=\frac{e^{-i\delta_{\hspace{-0.13ex} A}}}{i\sqrt
      2}\left[e^{-\halfi \delta_r}\ket{E^+} - e^{+\halfi
        \delta_r}\ket{E^-} \right]\!.
  \end{align}
\end{subequations}
where $\ket{E^\sigma}$ are the first stage scattering states at energy
$E$ with incoming flux in the direction $\sigma=\pm$. These
combinations are special in that they have real wavefunctions with the
asymptotic boundary conditions $\bracket{x}{E^S}\sim\frac{1}{\sqrt{\pi
    k}}\cos(kx-\half\delta_r\pm\delta_S)$ and
$\bracket{x}{E^A}\sim\frac{1}{\sqrt{\pi
    k}}\sin(kx-\half\delta_r\pm\delta_A)$ as $x\to\pm\infty$, where
the kinetic energy is $\half k^2$. These states are a generalization
of the treatment for a Hamiltonian with parity symmetry \cite{kahn}.
The asymptotics also hold for unequal leads by using their
respective wavenumbers. Time-reversal invariance implies that these
states' second stage coupling coefficients to the bound state
$v_E^{\SA}=\bra{\varphi}V^\mathrm{II}\ket{E^{\SA}}$ may be chosen to
be real.  Finally, the first-stage transmission coefficient is $
\bT^\mathrm{I}=\cos^2\big(\delta_S - \delta_A \big)$.

According to the Fano model \cite{fano1}, the first stage continuum is
decomposed into states that couple to the bound state and those that
don't. The coupled states acquire the Fano phase shift
$e^{2i\delta_E}$ due to the interaction with the bound state, where
$\cot\left(\delta_E\right)=-2(E-E_\varphi-\Delta_E)/\Gamma_E$, with
the usual definitions $\Gamma_E= 2\pi|v_{E}|^2$ and
$\Delta_E=\frac{1}{2\pi}\sP\int \!dE'\,\frac{\Gamma_{E'}}{E-E'}$. Here
$v_E$ is the radial coordinate of the point $(v_E^S,v_E^A)\in\RR^2$,
and we denote its angle as $\alpha$.

Finally, using these definitions, the total two-stage S-matrix is
\cite{MSP}
\begin{align}
  \label{eq:TwoStageSmatrixElements}
  S &= S^\mathrm{I}S^\mathrm{II} = e^{i(\delta_S+\delta_A+\delta_E)}\bpmat
  \tau && ie^{-i\delta_r}\rho^- \\
  ie^{+i\delta_r}\rho^+ && \tau \epmat
  \bmat \\, \emat \\
  \tau&=\cos^2 \!\! \alpha \cos(\delta_S-\delta_A +\delta_E) +
  \sin^2 \!\! \alpha \cos(\delta_S-\delta_A -\delta_E) \nonumber\\
  \rho^\pm&= \cos^2 \!\! \alpha \sin(\delta_S-\delta_A
  +\delta_E)
  + \sin^2 \!\! \alpha \sin(\delta_S-\delta_A -\delta_E) \nonumber\\
  &\pm 2i\sin\alpha\cos\alpha\sin\delta_E .\nonumber
\end{align}

If $v_E$ varies slowly enough as a function of $E$ then it is
reasonable to approximate $\Gamma_E$ and $\Delta_E$ using their values
at the bound state's energy $E_\varphi$ \cite{atomphoton}. These are
denoted as $\Gamma$ and $\Delta$.  In addition, we assume a constant
scattering background, i.e.  $S^\mathrm{I}$ and its associated
parameters are treated as constant, with values also taken at
$E_\varphi$.  This is a fair approximation when $\Gamma$ is much
smaller than the scale for which $S^\mathrm{I}$ changes appreciably.

Next, let's look at the Fano lineshapes. The ratio of the full
two-stage transmission probability to that of the first stage assumes
the almost familiar form
\begin{equation}
  \label{eq:TransmissionFanoProfile}
  \frac{\ \bT^{\phantom{\mathrm{I}}}}{\ \bT^\mathrm{I}}
  =\frac{\big(\epsilon+\qb\qi\big)^2}{1+\epsilon^2},
\end{equation}
with the usual reduced energy $\epsilon = -\cot \delta_E$ and the
novel factorization of the lineshape parameter in Eq.~\ref{eq:qfact} where
\begin{align*}
  \qb &= \tan \big(\delta_S-\delta_A \big)=\pm\sqrt{\bR^\mathrm{I}/\bT^\mathrm{I}}\\
  \qi &= \cos(2\alpha).
\end{align*}
In these terms, the actual two-stage transmission coefficients is
$\bT=(\epsilon+\qb\qi)^2/[(1+\epsilon^2)(1+\qb^2)]$.

The general form of the transmission Fano lineshape is described in
Fig.~\ref{fig:twistcone}(b). Note that a symmetric lineshape is
obtained not only when the background is zero ($\qb\to\pm\infty$
resulting in a Breit-Wigner Lorentzian) or unity ($\qb=0$ giving a
symmetric Lorentzian dip), but also when $\qi=0$. The latter occurs
when the bound state is coupled with equal strength to $\ket{E^S}$ and
$\ket{E^A}$.

\textit{Conical Intersections.}--- The lineshape asymmetry parameter
$q$ factors into a background contribution $\qb$ that solely depends
on $V^\mathrm{I}$, and a coupling interaction contribution $\qi$,
which may be controlled using $V^\mathrm{II}$. However, neither $\qb$
nor $\qi$ are gauge invariant when taken alone, only their product is.
This subtlety will be treated in detail in what follows.
Understanding how $V^\mathrm{II}$ controls $\qi$ is a simple affair
--- the decoupled first-stage surfaces $V^\mathrm{I}$ are not
affected, so that $\qb$ and the states $\ket{E^{\SA}}$ remain constant
regardless of the specific gauge used to define them.  On the other
hand, modifying $V^\mathrm{I}$ may affect both $\qb$ and $\qi$, an
effect that is predominant in the vicinity of background unity
transmission peaks.

Suppose that at a scattering energy $E=E_0$ the background
transmission reaches a unity peak, implying that
$\delta_S^0=\delta_A^0$. Here we accent values at the transmission
peak by a $0$ so that $S^\mathrm{I}_0=e^{i\theta_F^0}\1$. Assume
throughout that $\delta_S$ approaches $\delta_A$ with a leading order
linear in $E-E_0$. Since the background S-matrix is degenerate at this
energy, there are no preferred eigenvectors. However, we may choose to
use the states in Eqs.~\ref{eq:EigenStates} that evolve smoothly from
those obtained for values $E<E_0$, which we denote $\ket{E^{\SA}_0}$.
These limiting states are associated with an angle $\delta_r^0$, which
is just the limit of $\delta_r$ as $E_0$ is approached from below.
Note that by a suitable choice of gauge (i.e. choosing which point to
label $x=0$) it is possible to set $\delta_r^0=0$.

Now we modify $V^\mathrm{I}$ near the background transmission unity at
$E_0$, leading to three types of infinitesimal deformations of the
background S-matrix
\begin{displaymath}
  dS^\mathrm{I}= iS^\mathrm{I}_0\left[ \1 d\xi + \Sigma_r^0  \left(\sigma_x d\eta + \sigma_y d\zeta \right)  \right],
\end{displaymath}
where $\Sigma^0_r=\exp\big(-i\sigma_z \delta_r^0 \big)$. Note the
absence of a $\sigma_z$ \textit{generator} which breaks TRI.  The
$\xi$ deformation affects only $\theta_F$, keeping the degeneracy
intact, and it doesn't change the eigenvectors, nor does it change the
position of the transmission peak that still reaches unity. The $\eta$
deformation acts through the $\sigma_x$ generator. This removes the
degeneracy, but to leading order $\ket{E^{\SA}_0}$ are still its
eigenstates. It also shifts the position of the unity peak in the
transmission spectrum away from $E_0$.  Finally, the $\zeta$
deformations change the height of the peak whilst keeping its position.
More importantly, $\zeta$-deformations are orthogonal to the
$\eta$-deformations in the sense that the degeneracy is removed so
that to leading order the eigenvectors are proportional to
$\ket{E^S_0}\pm\ket{E^A_0}$.

The effect of these deformations on the first stage S-matrix may be
described as a conical intersection of the phase shifts $\delta_{\SA}$
in the $\eta\zeta$-plane, schematically depicted in
Fig.~\ref{fig:twistcone}(a). Here the actual eigenstates
$\ket{E^{\SA}}$ are defined according to the $\Delta\theta\ge0$ gauge,
i.e.  $\ket{E^S}$ is on the upper cone. The leading order behavior of
this state depends on the position in the parameter space. On one side
of the $\eta$ axis $\ket{E^S}\sim\ket{E^S_0}$, while on the other it
crosses over to $\ket{E^S}\sim\ket{E^A_0}$. Similarly, along the
$\zeta$ axis it crosses over from $\ket{E^S_0}+\ket{E^A_0}$ to
$\ket{E^S_0}-\ket{E^A_0}$. 

One might naturally ask what topological phase is associated with
these conical intersections. The answer is that $\delta_r$ shifts by
$2\pi$ for each cycle that encircles a vertex, and this serves as the
basis for adiabatic quantum swimming/pumping \cite{qswimmer}.

The interplay between $\qb$ and $\qi$ can now be understood in terms
of such conical intersections. The $\Delta\theta\ge0$ gauge choice
ensures that $\qb>0$ on the conical surface (equaling zero at the
vertex), so that up to appropriate parameterization dependent scale
factors, $\qb$ corresponds to the radial distance from the vertex in
parameter space.  In many cases it is also reasonable to assume that
the bound state's coupling to the $\ket{E^{\SA}_0}$ states remains
(relatively) constant near the intersection.  However, due to the
behavior of $\ket{E^{\SA}}$ that changes according to the position in
parameter space, the actual couplings to the bound state also change,
so that $\qi$ depends on the azimuthal angle's cosine.  The initial
coupling to the $\ket{E^{\SA}_0}$ states may be modified through
$V^\mathrm{II}$, and in effect this rotates the ``compass wheel'' in
Fig.~\ref{fig:twistcone}(a), i.e. $\qi\equiv 1$ points in a different
direction in the $\eta\zeta$-plane.

\textit{Quantum Billiard Example.}--- Recently, Fano resonances have
been studied extensively in connection with QBs, see e.g.
\cite{srotter,bird}. Usually, the QB is composed of an access lead
from which the electrons impinge upon the main billiard, typically the
part associated with a quantum dot, and an exit lead. The electron
scattering problem in QBs can be cast into a one-dimensional coupled
channel problem \cite{shachar,MSP}, where the channels are taken to be
the energies of the modes in the transverse direction, and the
inter-channel interactions are due to the non-adiabatic couplings
\cite{lenz,MSP}.

\begin{figure}[htbp]
  \centering
  \includegraphics[width=\setfigwidthratio{1}]{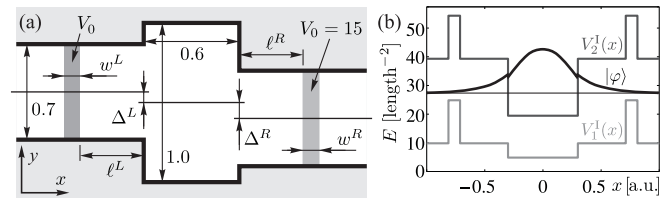}
  \caption{(a) Geometry of the quantum billiard with leads used to
    demonstrate control of Fano $q$-reversal. Length is measured in
    arbitrary units, while the energy scale is in units of
    [length$^{-2}$] (b) The first two adiabatic surfaces
    $V^\mathrm{I}_{1,2}(x)$ and the second surface bound state
    $\ket\varphi$.  }
  \label{fig:Billiard}
\end{figure}

Control of $q$ will be demonstrated on the QB depicted in
Fig.~\ref{fig:Billiard}(a). It consists of a rectangular cavity
connected to leads that have two potential barriers of constant
heights $V_0$, with widths $w^{\LR}$ and distances from the cavity
$\ell^{\LR}$ that may be varied to modify $V^\mathrm{I}$.  The
latter's first two adiabatic potential surfaces (channels) are
depicted in Fig.~\ref{fig:Billiard}(b). The leads may be offset by
$\Delta^{\LR}$ in order to control the non-adiabatic couplings
$V^\mathrm{II}$ resulting from the abrupt change in the local
transverse basis at the cavity edges. Although the formalism detailed
in Ref.~\onlinecite{MSP} is used to calculate the actual non-adiabatic
couplings, symmetry-based selection rules suffice for a qualitative
understanding.
 
The significance of a conical intersection is demonstrated near the
unity transmission peak at $E_0=27.35$ for a parity symmetric
$V^\mathrm{I}$ obtained by setting $w^\LR=0.1$ and $\ell^\LR=0.295$.
This is convenient since the symmetry ensures that the states
$\ket{E^\SA_0}$ have respective even and odd parity in the
$\delta_r\equiv0$ gauge.  The background $V^\mathrm{I}$ is modified to
control the lineshape parameter $q$ near the conical intersection.
Taking $\Delta^\LR=0.01$ ensures that $V^\mathrm{II}$ couples
$\ket\varphi$ exclusively to $\ket{E^S_0}$ by symmetry. We choose
$\eta=\ell^L=\ell^R$, which preserves parity, and $\zeta$-deformations
that break this symmetry.

\begin{figure}[htbp]
  \centering
  \includegraphics[width=\setfigwidthratio{0.95}]{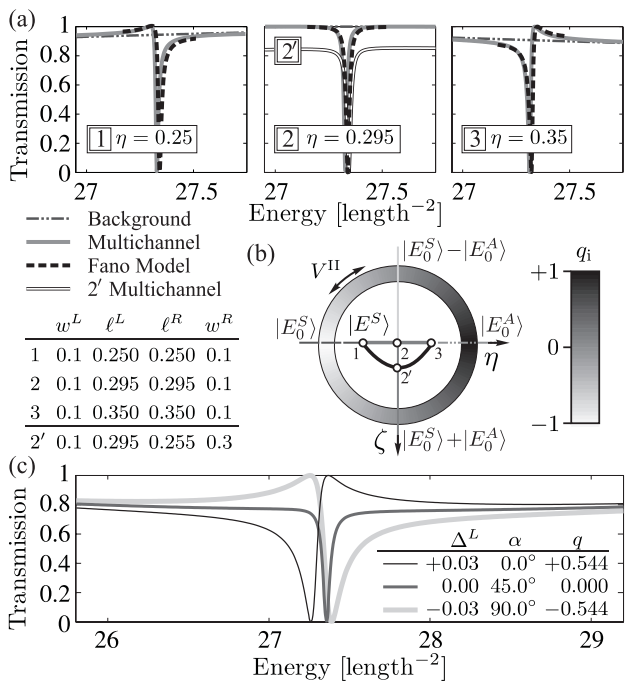}
  \caption{ (a) Multichannel and analytical lineshapes obtained for
    various $V^\mathrm{I}$ parameters (see table). (b) These
    correspond to 2 paths in parameter space where $q$ reversal
    occurs. The overall lineshape parameter vanishes at the
    intermediate points: $\qb=0$ at point 2 while $\qi=0$ at $2'$. (c)
    Variation of the non-adiabatic couplings $V^\mathrm{II}$, gives
    $q$-reversal due to the change in the coupling angle $\alpha$, and
    hence $\qi$. The table summarizes analytical values calculated
    from the model.  }
  \label{fig:Symmetric_q_b-reversal}
\end{figure}

Figure~\ref{fig:Symmetric_q_b-reversal}(a) depicts the Fano profiles
resulting from the numerically exact multichannel calculation
\cite{sheng2}, our analytical model, and the background
contributions for several sets of parameters, see table. In all cases
note the excellent fit between the actual multichannel calculations
and the lineshapes predicted by our model.  This series of
calculations corresponds to two $q$-reversal paths in parameter space
depicted in Fig.~\ref{fig:Symmetric_q_b-reversal}(b). The first path
along the $\eta$ axis goes from point 1, with $q<0$, to 3 ($q>0$) by
passing exactly through the background unity transmission peak at 2,
where $q_\mathrm{b}=0$. Note that for point 1 the transmission peak's
energy has moved to the right of the resonance, while the opposite
happens at point 3. The second path goes from 1 to 3 through $2'$
which breaks parity symmetry. This point has a sub-unity background
transmission peak, corresponding to a pure $\zeta$-deformation. As 
expected from the conical intersection, at this point in parameter
space the states $\ket{E^\SA}$ are similar to
$\ket{E^S_0}\pm\ket{E^S_A}$. This implies equal coupling to the bound
state, so that $\qi=0$, giving a symmetric dip with the novelty of a
less than unity background.

Control of $q$ may also be affected through $V^\mathrm{II}$, which
directly controls $\qi$, whilst $\qb>0$ remains constant. In the
language of conical intersections the $V^\mathrm{I}$ parameters remain
constant, so that $q$-reversal is achieved through a rotation of the
``compass wheel''. This type of control works irrespective of the
position in parameter space relative to conical intersections. For
simplicity, again we choose a parity symmetric $V^\mathrm{I}$, this
time with $\ell^{\LR}=0.4125$ and $w^{\LR}=0.1$ so that $\qb=0.544$
throughout.  Now we vary $\Delta^L=-0.03\mapsto 0.03$, while holding
$\Delta^R=0.03$. The results of the numerically exact multichannel
calculations are depicted in Fig.~\ref{fig:Symmetric_q_b-reversal}(c).
The results are easily explained in terms of symmetry if we use the
$\delta_r\equiv0$ gauge. Since the second channel bound state has even
parity, it is clear that for $\Delta^L=-0.03$ it couples exclusively
to $\ket{E^A}$, while for $\Delta^L=+0.03$ it is coupled only to
$\ket{E^S}$. Equal coupling, with $q_i=0$, is achieved for
$\Delta^L = 0.00$, yielding a sub-unity symmetric dip.

In summary, we have presented an analytical model for Fano resonances
in coupled two-channel systems, where we found that the Fano lineshape
parameter $q$ factors into background and interaction contributions.
The model give accurate predictions for the actual transmission
lineshapes. Moreover, it also provides insight to the relation between
conical intersections of the background phase shifts, the coupling
interaction, and the overall lineshape parameter $q$. This allows full
control of $q$, which was demonstrated for a quantum billiard example.

The authors acknowledge useful discussions with N. H. Lindner and N. Moiseyev. This work was funded in part by the fund for promotion of research at the Technion.



\end{document}